\documentclass[a4paper]{jpconf}
\usepackage{graphicx}
\begin{document}
\title{Open-charm production as a function of charged particle multiplicity in pp collisions at $\sqrt{s}$=7 TeV with ALICE}

\author{Renu Bala, for the ALICE Collaboration}

\address{Department of Physics, University of Jammu}

\ead{renu.bala@cern.ch}

\begin{abstract}
Heavy quarks (charm and beauty) are an effective tool to investigate
the properties of the Quark-Gluon Plasma created in heavy-ion collisions as
they are produced in initial hard scattering processes and as they
experience all the stages of the medium evolution. The measurement of
heavy-flavour production cross sections in pp collisions at the LHC,
besides providing a reference for heavy-ion studies, allows one to
test perturbative QCD calculations. A brief review of ALICE results on
the production of heavy-flavoured hadrons measured from fully 
reconstructed hadronic decay topologies in pp collisions at $\rm
\sqrt{s}$ =  7 TeV is  presented.
Furthermore, heavy-flavour production was also studied as a function
of the particle multiplicity in pp collisions. This could provide
insight into multi-parton scatterings.  A measurement of the inclusive J/$\rm \psi$ yield as a function of
the charged-particle pseudorapidity density was performed by the ALICE
Collaboration at the LHC in pp collisions at $\rm \sqrt{s}$ = 7
TeV. An increase of the J/$\rm \psi$ yield with increasing
multiplicity was observed. In this context, the study of the yield of
D mesons as a function of the charged-particle multiplicity could
provide a deeper insight into charm-quark production in pp
collisions. We will present the first results obtained for prompt $\rm
D^{0}$, $\rm D^{+}$, and $\rm D^{*+}$
mesons using hadronic decay channels at midrapidity in pp collisions
at $\rm \sqrt{s}$=7 TeV as a function of the charged-particle multiplicity. The prompt D-meson
yields as a function of multiplicity are measured in different $p_{\rm
  T}$ intervals. These yields will be compared to the results obtained for inclusive and non-prompt
J/$\rm \psi$.
\end{abstract}

\section{Introduction}
Heavy quarks are unique probes to study the Quark-Gluon Plasma
produced in heavy ion collisions at the LHC. Due to their large
masses, they are produced predominantly in hard parton scattering processes, during
the initial stages of the collision. Therefore, they experience the entire evolution of the medium created in the collision and can act as probes of its properties. The measurement of  heavy-quark production in pp
collisions besides providing a necessary reference for the study of
medium effects in Pb-Pb collisions, serves as  a precision test for
perturbative Quantum Chromodynamics (pQCD). In addition, the study of heavy-quark production as a
function of charged particle multiplicity  could provide  insight into
multiple hard parton scattering. This allows one to
investigate the interplay between hard and soft QCD processes and to
study the role of multi-parton interactions. It is interesting to
remark that the highest multiplicities reached in pp collisions at 7 TeV are similar to those observed in Cu-Cu collisions at RHIC
energies.
\section{Open charm measurement with ALICE} ALICE (A Large Ion Collider
Experiment)  is  the LHC experiment dedicated to heavy ion
studies. ALICE  \cite{Alice} consists of two parts: a 
barrel at central rapidity and a muon spectrometer at forward
rapidity. For the present analysis, we have used the information
from a subset of the central barrel detectors, namely the Inner
Tracking System (ITS), the Time Projection Chamber (TPC) and the Time Of
Flight detector (TOF) for charged particle tracking and
identification, the T0 for time zero measurement and the VZERO
scintillator for triggering. The two tracking detectors, the ITS and
the TPC, allow the reconstruction of charged-particle tracks in the
pseudorapidity range -0.9 $< \rm \eta <$ 0.9 with a momentum resolution
better than 2$\%$ for $ p_{\rm T} < $ 20 GeV/$c$ and they provide particle
identification via a d$E$/d$x$ measurement.  The ITS, in particular, is a key
detector for open heavy flavour studies because it allows us  to measure
the track impact parameter (i.e. the distance of closest approach of
the track to the primary vertex) with a resolution better than 75 $\rm
\mu$m for $ p_{\rm T} >$ 1 GeV/$c$ thus providing the capability to
detect secondary vertices originating from heavy-flavour hadron decays. The TOF detector provides particle identification by time of flight measurement. 

The results presented here are obtained from pp  data recorded in 2010 at
$\rm \sqrt{s}$= 7 TeV. The analyzed sample was collected with a
minimum bias trigger condition requiring at least one hit in the
Silicon Pixel Detector (SPD) (equipping the two innermost layers of
the ITS), or a signal in at least one of the two VZERO detectors, which
cover the pseudorapidity ranges $\rm -3.7 < \eta < -1.7$ and $ \rm 2.8
< \eta < 5.1$.

$\rm D^{0}, D^{+},
D^{*+}$ and $\rm D_{s}^{+}$  mesons are reconstructed in the central
rapidity region  from their hadronic decay channels  $\rm D^{0} \rightarrow K^{-} \pi^+, D^{+}
\rightarrow K^{-} \pi^{+} \pi^{+}, D^{*+} \rightarrow  D^{0} \pi^{+}
\rightarrow K^{-} \pi^{+}
\pi^{+}$ and $\rm D_{s}^{+} \rightarrow \phi \pi^{+} \rightarrow K^{+}
K^{-} \pi^{+}$ (and charge conjugates). The D-meson yield is measured
with an invariant mass analysis of fully reconstructed decay topologies displaced from the
interaction vertex selected by requiring a large decay length and a good alignment between
the reconstructed D-meson momentum vector and flight line. The identification
of charged kaons in the TPC and TOF detectors helps to further
reduce the background at low $ p_{\rm T}$.

\section{D-meson production cross section measurement}
\begin{figure}[h]	 
\includegraphics[width=13pc,height=13pc]{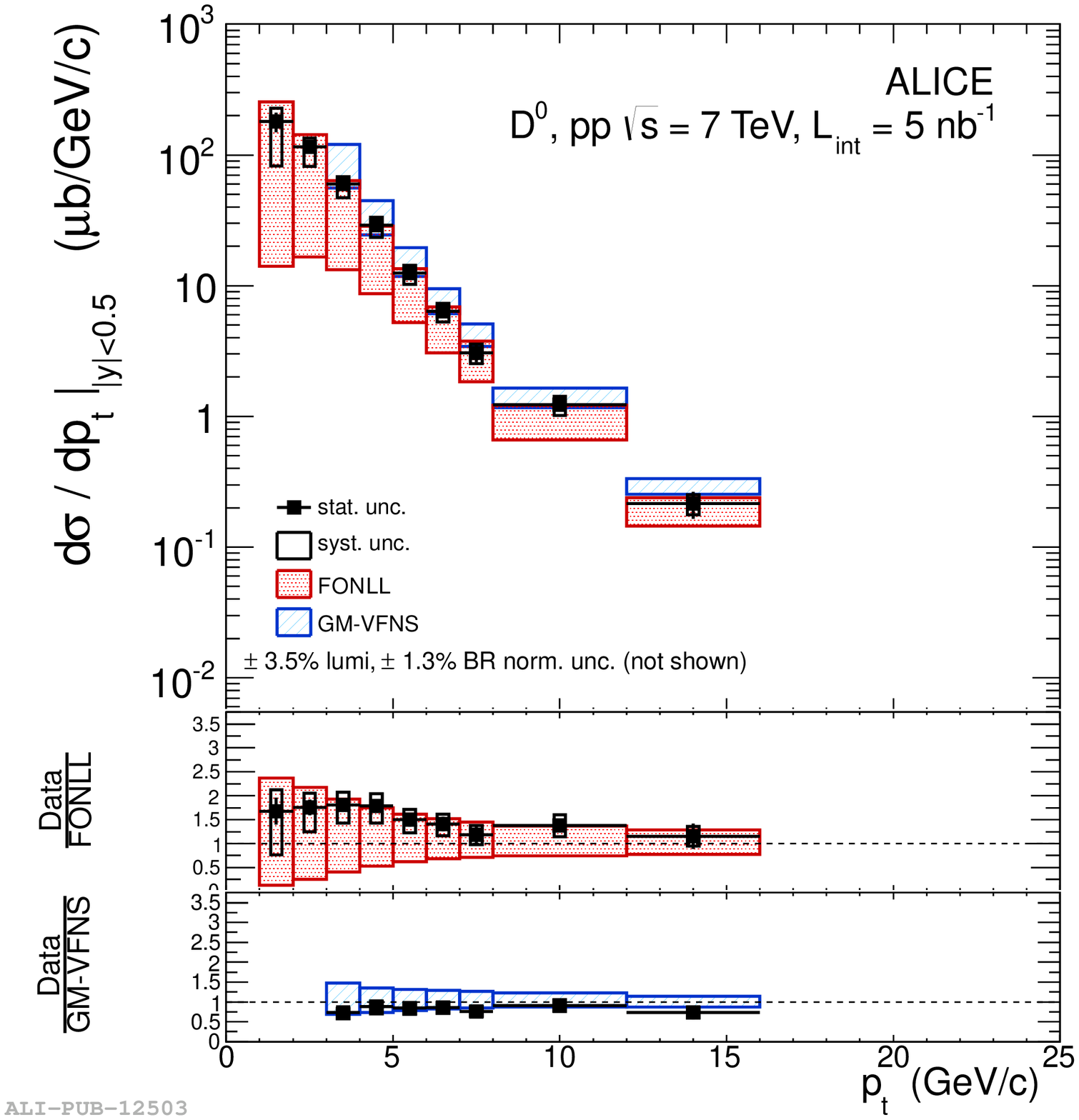}
\includegraphics[width=13pc,height=13pc]{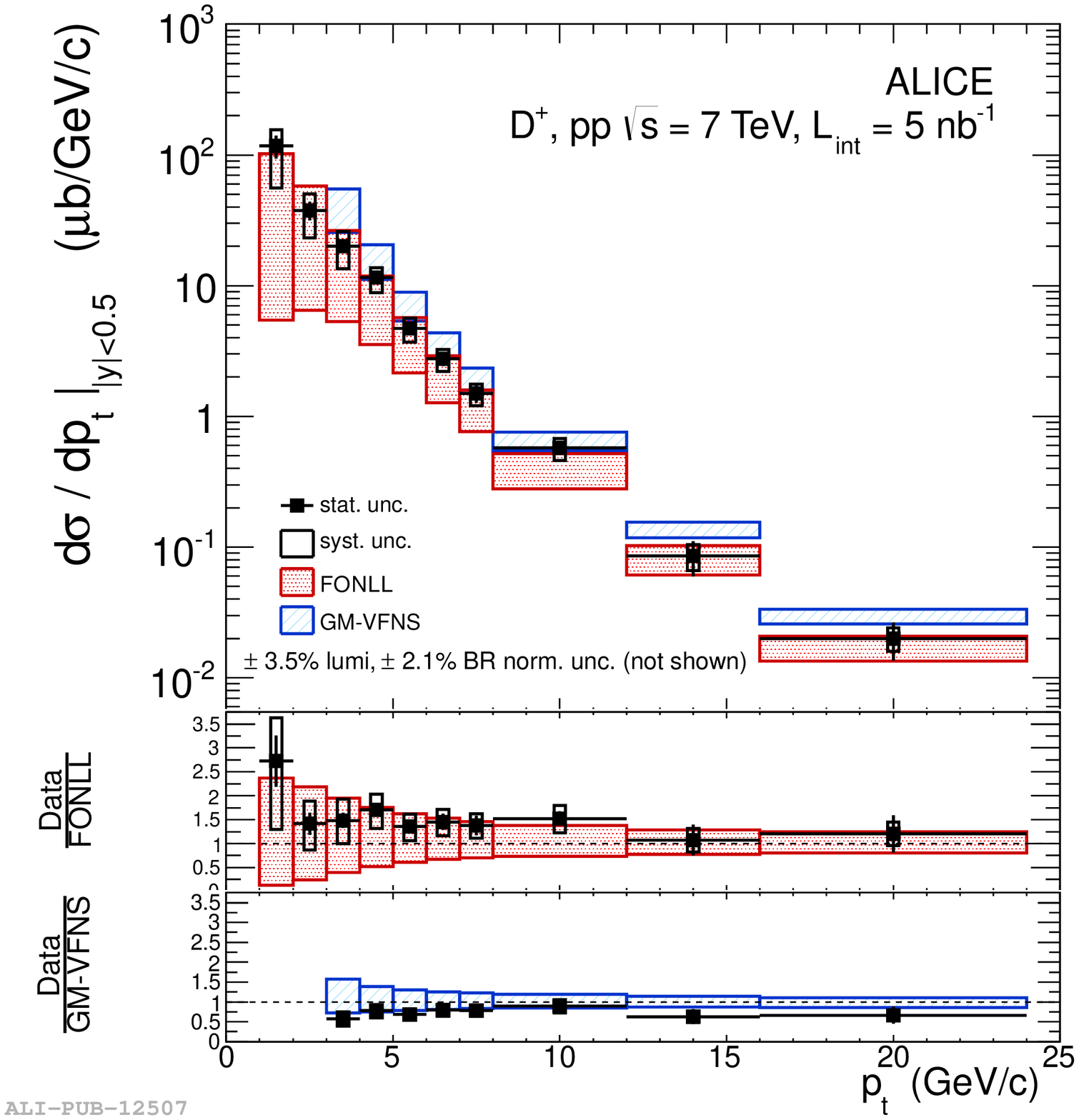}
\includegraphics[width=13pc,height=13pc]{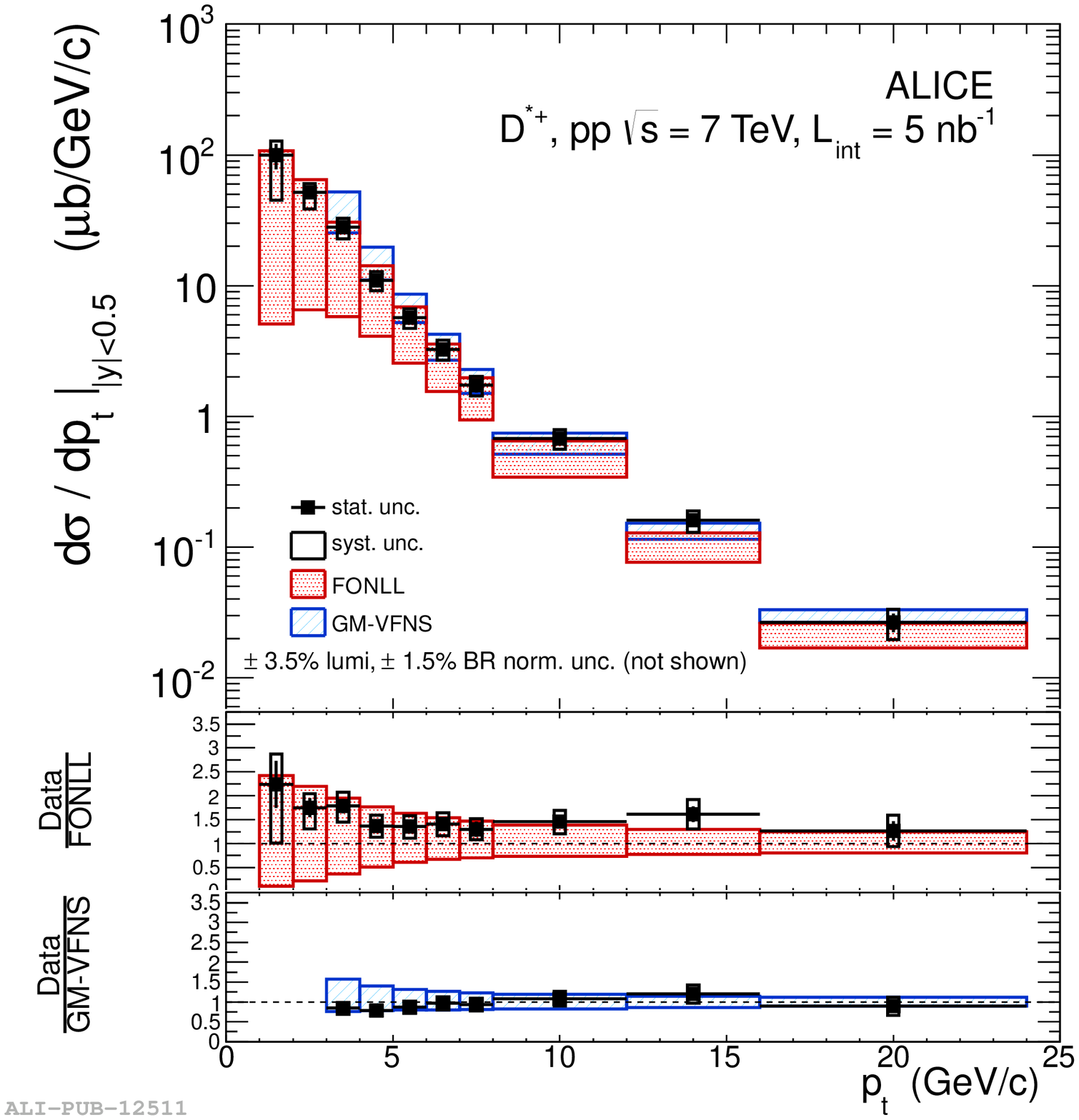}
\caption{  $ p_{\rm T}$ differential cross section  for $\rm D^{0}$ (left panel),
$\rm D^{+}$ (middle panel) and $\rm D^{*+}$ (right panel) in pp collisions
at $\rm \sqrt{s}$=7
TeV, compared to FONLL \cite{cacciari} and
GM-VFNS \cite{GM} theoretical predictions.}
\label{sigma}

\end{figure}

The differential cross section has been measured for prompt $\rm D^{0}, D^{+},
D^{*+}$ and $\rm D_{s}^{+}$ mesons in pp collisions at $\rm \sqrt{s}$ = 7
TeV \cite{Dppaper,Dspaper}.  To obtain the cross section for prompt
D-mesons, the contribution due to  beauty hadron decays, which is of
about 10-15$\%$, was evaluated based on Fixed Order with Next-to-Leading-Log resummation
(FONLL) pQCD calculations \cite{cacciari} and subtracted from  the
inclusive D-meson yield.  The
resulting cross sections are compared to FONLL  and General Mass Variable Flavour
Number Scheme (GM-VFNS) \cite{GM} pQCD calculations.
 Figure ~\ref{sigma} shows the $p_{\rm T}$-differential cross section for prompt $\rm D^{0}$, $\rm D^{+}$ and
$\rm D^{*+}$ mesons at $\rm \sqrt{s}$ = 7 TeV.  The data are well described by pQCD calculation: at the
upper limit of the uncertainty band when compared to  FONLL and at
the  lower limit when compared to GM-VFNS 
predictions.  Recently, it was reported  that data are also well
described  by the $\rm k_{T}$ factorization approach \cite {kTfact}.

\section{D-meson yields as a function of charged-particle multiplicity}

\begin{figure}[h]
\begin{center}
\includegraphics[width=16pc,height=15pc]{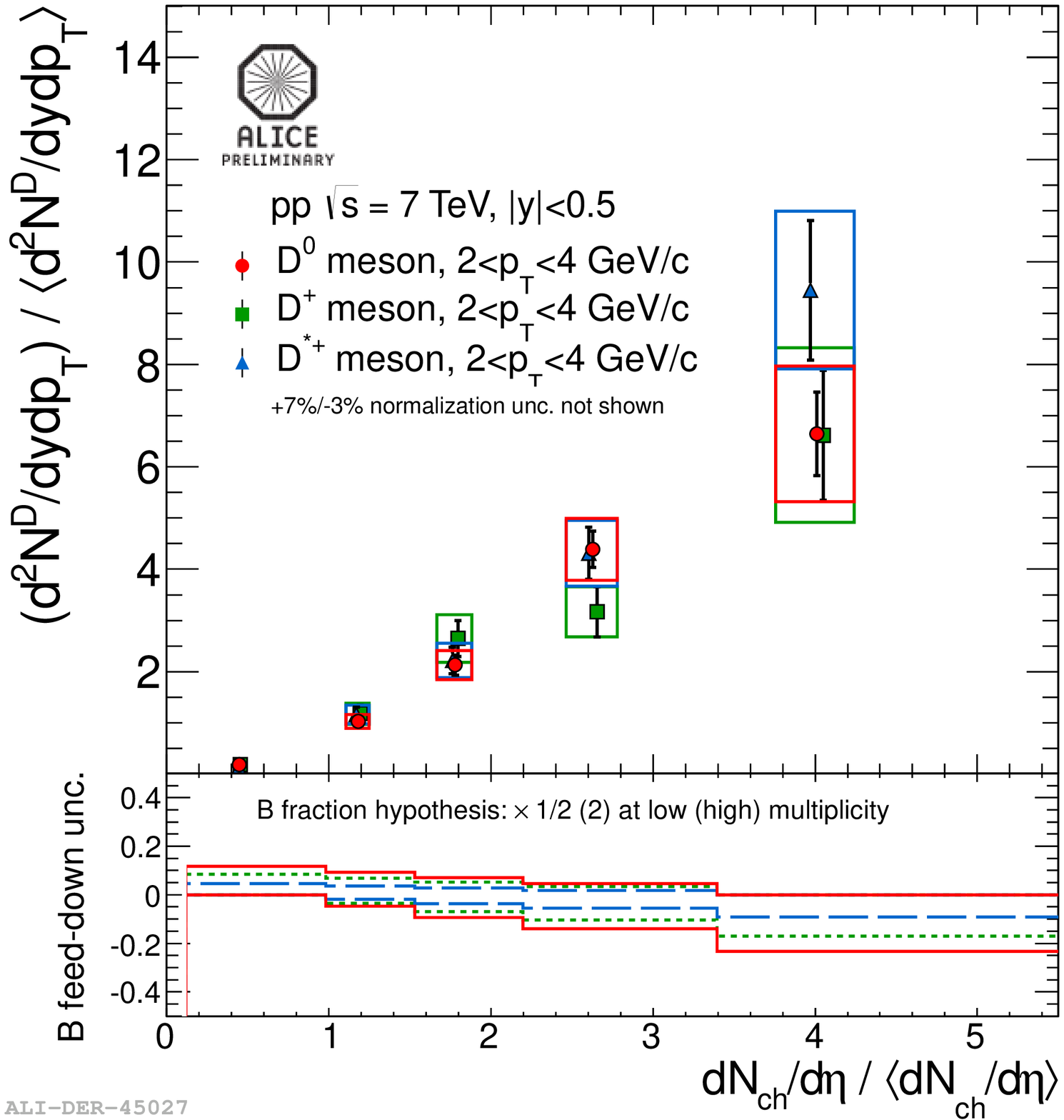}
\includegraphics[width=16pc,height=15pc]{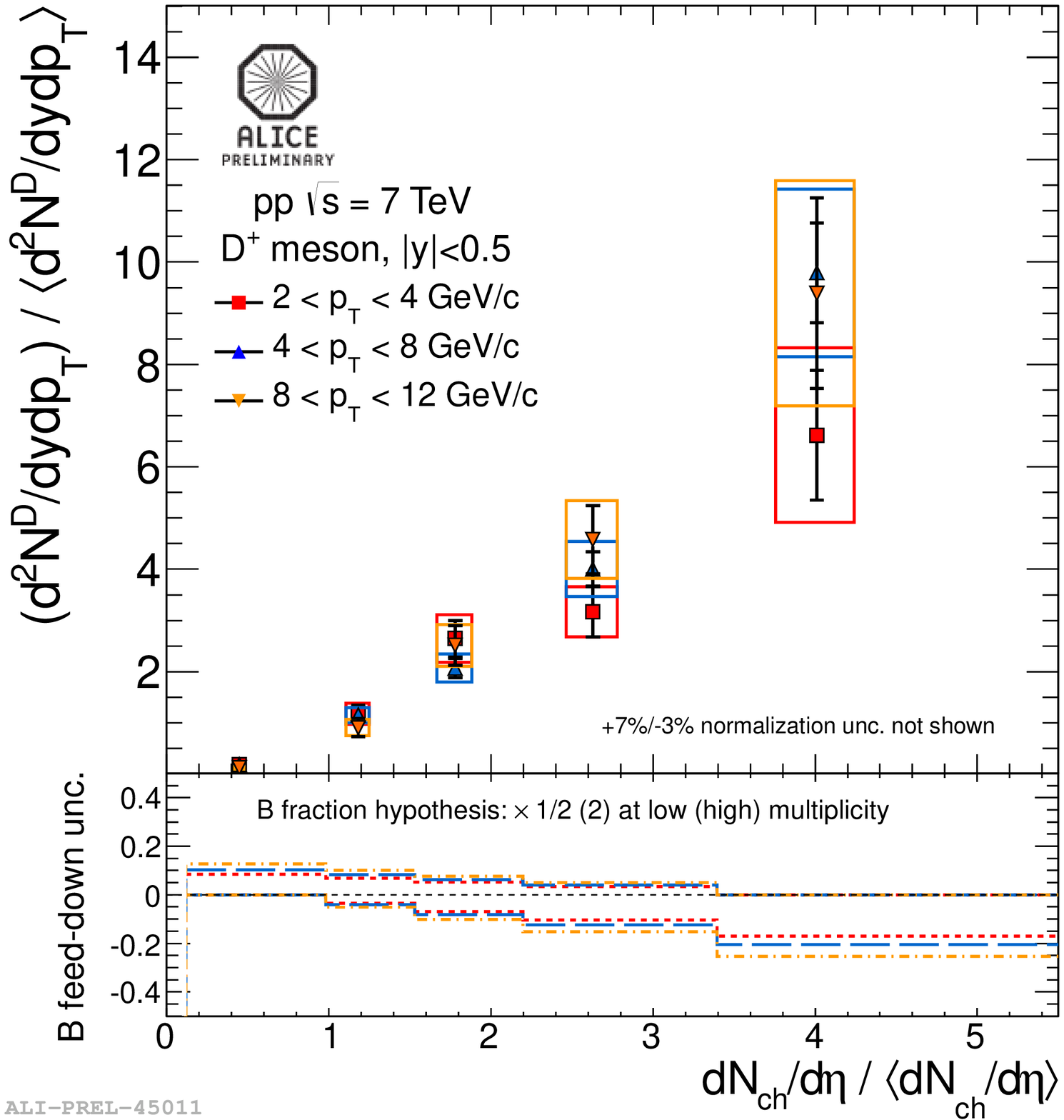}

\caption{ Left: Relative $\rm D^{0}$ , $\rm  D^{+}$ and $\rm D^{*+}$ yields
  in 2 $< p_{\rm T} <$ 4 GeV/$c$ as a function of the relative multiplicity
of charged particles produced in pp collision at $\rm \sqrt{s}$ = 7
TeV. Right: Relative $\rm  D^{+}$  yields in three $p_{\rm T}$
intervals as a function of the relative multiplicity
of charged particles produced in pp collision at $\rm \sqrt{s}$ = 7 TeV}
\label{Relmult}
\end{center}
\end{figure}

The production of D mesons is also studied in several $p_{\rm T}$
intervals as a function of the multiplicity  of charged particle
generated in the collision. The multiplicity estimator used in this analysis is the number of SPD
tracklets (combination of two hits in the two layers of SPD, $\rm N_{trk}$) in $ \rm |\eta| <$ 1.0. Since the pseudorapidity coverage
of the SPD
changes with the z position of the interaction vertex, a correction to the
measured $\rm N_{trk}$ is applied event-by-event. Using simulated events, it is verified
that $\rm N_{trk}$ is proportional to d$\rm N_{ch}$/d$\rm \eta$ (pseudorapidity density of primary charged particles produced in the collision). The D-meson yields are extracted in five intervals of multiplicity.
 The left panel of  
figure ~\ref{Relmult} shows the  yield of $\rm D^{0}, D^{+}$
and $\rm D^{*+}$ in a given multiplicity interval divided by the value
integrated over multiplicity for 2 $<p_{\rm T}<$ 4 GeV/$c$ as a function of
charged-particle multiplicity, expressed as the ratio between the $\rm
dN_{ch}$/d$\rm \eta$ in the considered
multiplicity interval and the $\rm <dN_{ch}/d\eta>$ for pp collisions
at 7 TeV.  The horizontal size of the error boxes represents
the systematic uncertainty in the $\rm
dN_{ch}/d\eta/<dN_{ch}/d\eta>$. The vertical size of the error boxes
reflects all  uncertainties on the relative yield except for the feed
down contribution. The latter  uncertainty is shown in the
lower panel of each figure. To account for a possible difference in the
multiplicity distribution of events with c$\rm \bar{c}$ and b$\rm
\bar{b}$  production, the systematic uncertainty was estimated by allowing the B/D ratio predicted by FONLL to vary by a factor 2 up(down) at high(low) multiplicity.
The results for  $\rm D^{0}, D^{+}$ and $\rm D^{*+}$
are compatible within the statistical and systematic
uncertainties and show an increase of the D-meson yields with the
charged-particle multiplicity of the event. The same  trends are
observed in all the measured  $ p_{\rm T}$ intervals as shown in the
right panel of figure ~\ref{Relmult}. Within the current
uncertainties, no $p_{\rm T}$ dependence
is observed.

\begin{figure}[ht]
\begin{center}
\includegraphics[width=16pc,height=15pc]{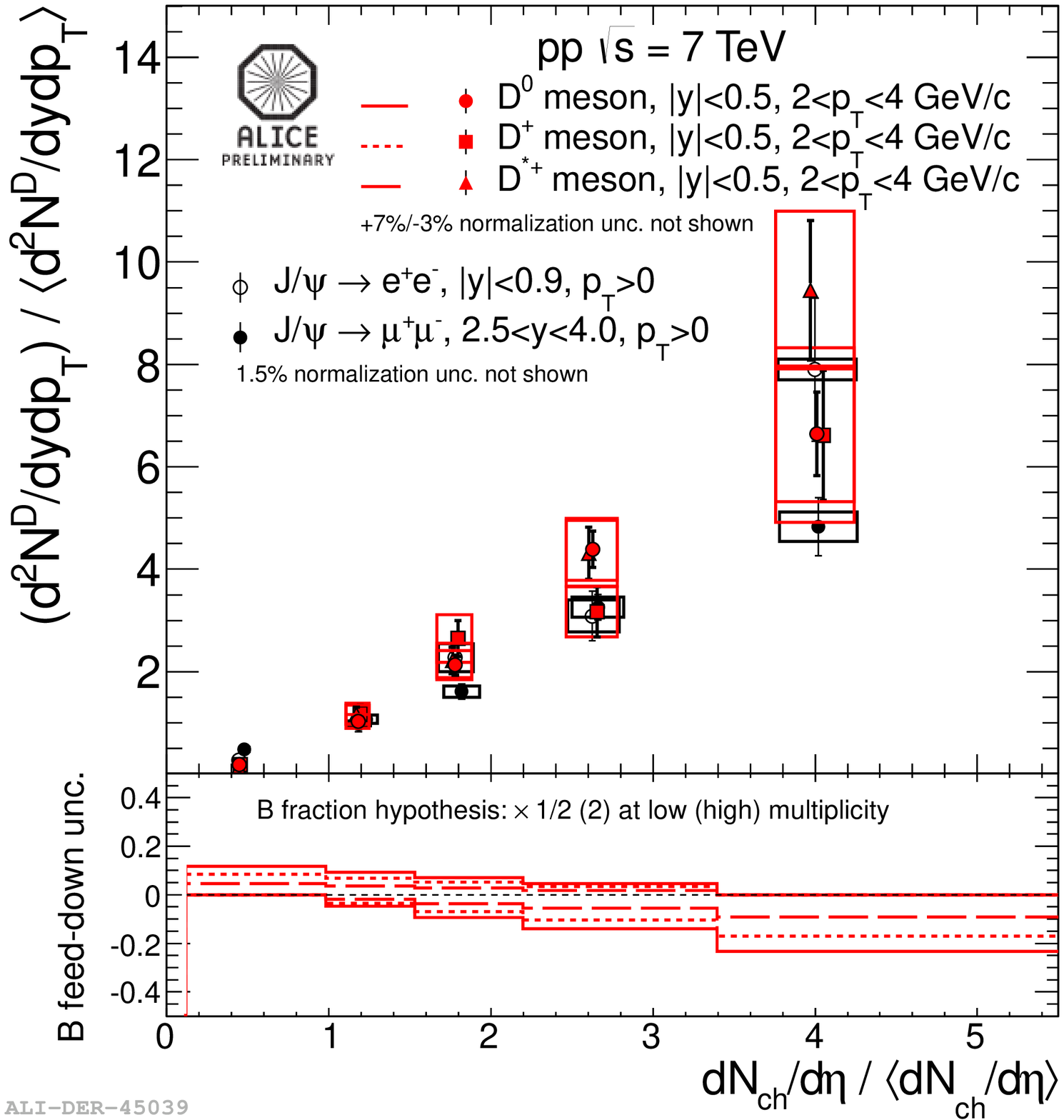}
\includegraphics[width=16pc,height=15pc]{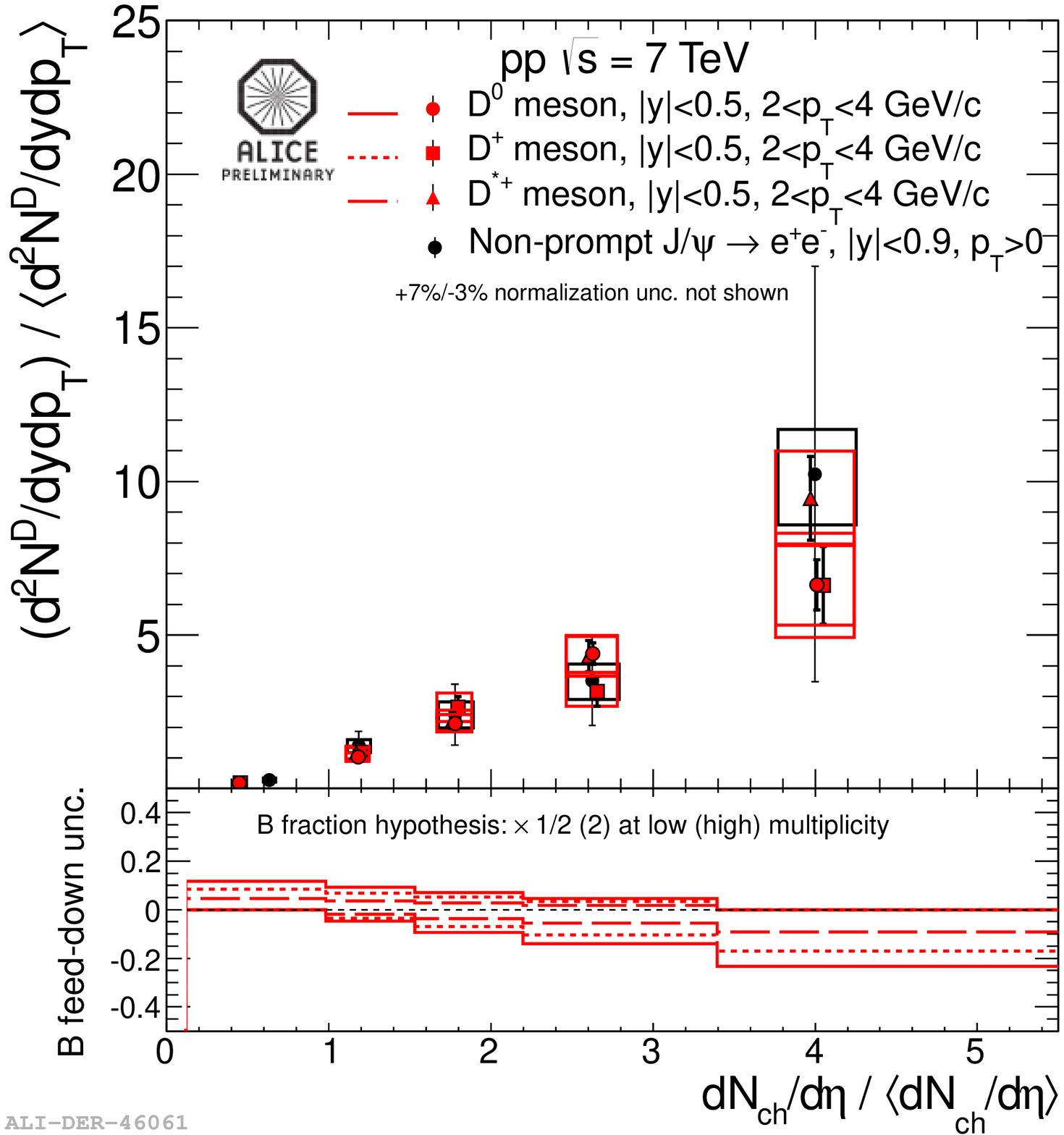}
\caption{Left: Relative yield of $\rm D^{0}$, $\rm D^{+}$, $\rm D^{*+}$ for
  2 $ < p_ {\rm T} <$ 4 GeV/$c$ and J/$\rm \psi$ for $ p_{\rm T} >$ 0
  as function of the event charged particle multiplicity. Right:  Relative yield of $\rm D^{0}$, $\rm D^{+}$, $\rm D^{*+}$ for
  2 $< p_{\rm T} <$ 4 GeV/$c$ and non prompt J/$\rm \psi$ 
  as function of the event charged particle multiplicity.}
\label{RelmultJpsi}
\end{center}
\end{figure}

The left panel of figure ~\ref{RelmultJpsi}  shows the  comparison of
D meson and  inclusive J/$\rm \psi$ yields as a function of multiplicity \cite{jpsipaper}. The J/$\rm
\psi$ mesons  were measured in  two rapidity intervals, namely
$|y| <$ 0.9 and 2.5 $< y < $ 4.0,  whereas the D-meson rapidity and
$p_{\rm T}$ intervals are $|y| <$ 0.5 and 2 $ < p_{\rm T} <$ 4 GeV/$c$ respectively. Both open
and hidden charm show similar behavior as a function of charged-particle multiplicity. Also the comparison with J/$\rm
\psi$ from B-hadron decays in $|y| <$ 0.9 (figure ~\ref{RelmultJpsi} right)
shows a similar increase of yield.

\section{Conclusions}
 ALICE detector provides excellent tracking, vertexing and particle
identification to allow the measurement of  charmed mesons via their
hadronic decays, over a wide range of transverse momentum. The prompt D-meson production cross section is well described by NLO pQCD
calculations. Furthermore, the D-meson yields was studied as a
function of the charged particle multiplicity in pp collisions at $\rm \sqrt{s}$=7 TeV. An
increase of the D-meson yield as a function of multiplicity is observed. The comparison with the results for  prompt and non-prompt
J/$\psi$ shows a similar increase as a function of charged-particle
multiplicity. These results may indicate that either D-meson production in pp collisions is connected with a strong hadronic activity, or multi-parton interactions also affect the hard momentum scales relevant for charm quark production.

\end{document}